\documentclass[runningheads]{llncs}
\usepackage{pgfplots} 
\pgfplotsset{compat=1.16}

\usepackage[caption=false]{subfig}
\usepackage[colorlinks=true,linkcolor=amaranth]{hyperref}

\usepackage{siunitx}

\usepackage{xfrac}

\usepackage{booktabs}
\usepackage{xcolor} 
\definecolor{azure}{rgb}{0.0, 0.5, 1.0}
\definecolor{darkgreen}{rgb}{0.0, 0.5, 0.0}
\definecolor{amaranth}{rgb}{0.9, 0.17, 0.31}
\definecolor{cadetgrey}{rgb}{0.57, 0.64, 0.69}
\definecolor{aureolin}{rgb}{0.99, 0.93, 0.0}

\usepackage{amssymb}

\usepackage{listings}

\newcommand{\python}[1]{\lstinline[language=Python]{#1}}
\newcommand{\swift}[1]{\lstinline[language=Swift,morekeywords={await}]{#1}}

\lstdefinelanguage{rust}
{
    keywords={
    true,false,
    unsafe,async,await,move,
    use,pub,crate,super,self,mod,
    struct,enum,fn,const,static,let,mut,ref,type,impl,dyn,trait,where,as,
    break,continue,if,else,while,for,loop,match,return,yield,in
    },
    ndkeywords={
    bool,u8,u16,u32,u64,u128,i8,i16,i32,i64,i128,char,str,Self,Option,Some,None,
    Result,Ok,Err,String,Box,Vec,Rc,Arc,Mutex,Cell,RefCell,HashMap,BTreeMap,macro_rules
    }
}
\newcommand{\rust}[1]{\lstinline[language=Rust]{#1}}
\newcommand{\go}[1]{\lstinline[language=Go]{#1}}

\usepackage[T1]{fontenc}
\usepackage{graphicx}
\usepackage{amsmath}

\begin{document}
\title{Benchmarking the Parallel 1D Heat Equation Solver in Chapel, Charm\texttt{++}, C\texttt{++}, HPX, Go, Julia, Python, Rust, Swift, and Java }
%
\titlerunning{Benchmarking the Parallel 1D Heat Equation Solver in various languages}
%
\author{
Patrick Diehl\inst{1,2}\orcidID{0000-0003-3922-8419} \and Max Morris\inst{1}  \and Steven R. Brandt\inst{1} \and Nikunj Gupta\inst{4} \and  Hartmut Kaiser\inst{1,3}\orcidID{0000-0002-8712-2806}}
\authorrunning{P. Diehl et al.}
%
\institute{Center of Computation \& Technology, Louisiana State University
\email{\{pdiehl,mmorris,sbrandt,hkaiser\}@cct.lsu.edu} \\
\and Department of Physics and Astronomy, Louisiana State University 
\and Department of Computer Science, Louisiana State University
\and Amazon LLC (Work done before joining Amazon)
\email{nknikunj@amazon.com}
}
\maketitle              
\begin{abstract}
Many scientific high performance codes that simulate \emph{e.g.}\ black holes, coastal waves, climate and weather, etc. rely on block-structured meshes and use finite differencing methods to solve the appropriate systems of differential equations iteratively. This paper investigates implementations of a straightforward simulation of this type using various programming systems and languages. We focus on a shared memory, parallelized algorithm that simulates a 1D heat diffusion using asynchronous queues for the ghost zone exchange. We discuss the advantages of the various platforms and explore the performance of this model code on different computing architectures: Intel, AMD, and ARM64FX. As a result, Python was the slowest of the set we compared. Java, Go, Swift, and Julia were the intermediate performers. The higher performing platforms were C\texttt{++}, Rust, Chapel, Charm\texttt{++}, and HPX.

\keywords{Asynchronous programming  \and Concurrency \and Julia \and Chapel \and Rust \and Go \and Charm\texttt{++} \and HPX \and Swift \and Java \and C\texttt{++}.}
\end{abstract}
\section{Introduction}
Several languages and libraries are emerging to provide better support for parallel programming, particularly asynchronous programming.
This work explores some of the frameworks available for shared memory parallelism.

Asynchronous execution can be implemented in various ways, \emph{e.g.}\ via coroutines, senders/receivers, futures, or queues. Table~\ref{tab:approaches:overview} provides an overview of the features for parallelism supported by the frameworks investigated.
This work implements a 1D stencil-based heat equation solver as a model problem, using asynchronous queues to exchange ghost zone cells.

We implement this model code using several different platforms: \textit{(1)} Chapel~\cite{chamberlain2015chapel} by HPE; \textit{(2)} Julia~\cite{bezanson2017julia} by Julia Inc, Go by Google; \textit{(3)} Swift by Apple; \textit{(4)} Charm\texttt{++}~\cite{kale1993charm++} by Charmworks; \textit{(5)} Standard C\texttt{++}; \textit{(6)} HPX~\cite{kaiser2020hpx} developed primarily at LSU; \textit{(7)} Python developed by open source communities; and \textit{(8)} Rust driven by Mozilla.

Previously, a comparison of Parallel Research Kernels for MPI, MPI+OpenMP, UPC, Charm\texttt{++}, and Grappa was made in 2016 in~\cite{10.1007/978-3-319-41321-1_17}. A comparison of \textit{programmability, performance, and mutability} of Charm\texttt{++}, Legion, and Unitah was made in 2015~\cite{bennett2015asc}. A comparison of Julia, Python\textbackslash Numba, and Kokkos was made in~\cite{godoy2023evaluating}.

This paper will compare the benefits and challenges of a much broader set of languages. We faced several challenges in implementing the appropriate solver. How much work did it take to achieve the performance we finally accepted? Next, we compare the performance of the languages on Intel, AMD, and A64FX CPUs. Finally, we discuss the trade-offs of using each of the various platforms.

\begin{table}[tb]
    \centering
    \caption{Overview of the programming languages: \textit{(1)} the parallelism approaches they provide, \textit{(2)} supported operating systems, \textit{(3)} the license, and \textit{(4)} reference. \textcolor{black}{The C++ 17 standard was used as a base. The symbol $\sim$ indicates that Charm\texttt{++} uses internal coroutines, but these are not accessible to the user, and Chapel solely provides a parallel for loop.} }
    \label{tab:approaches:overview}
    \begin{tabular}{l|ccc|ccc|lc}\toprule
     Approach    & Async  & Coroutine & Parallel algorithm & Win & Linux & Mac & Licence & Ref \\\midrule
    C\texttt{++}\ 17   & \checkmark & \checkmark & \checkmark &  \checkmark & \checkmark & \checkmark & GNU  & -- \\
     Java & \checkmark & \text{\sffamily X} &  \text{\sffamily X} & \checkmark & \checkmark & \checkmark & GNU & \cite{arnold2005java}\\
     Swift & \checkmark & \checkmark & \text{\sffamily X} & \checkmark & \checkmark & \checkmark & Apache & --  \\
    Chapel  & \checkmark  &  \text{\sffamily X} & $\sim$ & \checkmark & \checkmark & \checkmark & Apache  &   \cite{chamberlain2007parallel} \\
    Charm\texttt{++} & \checkmark  & $\sim$ & \text{\sffamily X}  & \checkmark & \checkmark & \checkmark & Own & \cite{kale1993charm++} \\
     HPX & \checkmark & \checkmark & \checkmark & \checkmark & \checkmark & \checkmark & Boost &   \cite{kaiser2020hpx} \\
     Go & \text{\sffamily X} & \text{\sffamily X} & \text{\sffamily X} & \checkmark & \checkmark & \checkmark & BSD  & --  \\
     Python & \checkmark & \checkmark & \text{\sffamily X} & \checkmark & \checkmark & \checkmark & BSD & \cite{10.5555/1593511}  \\
Julia & \checkmark  & \checkmark & \text{\sffamily X} & \checkmark & \checkmark & \checkmark &  MIT & \cite{bezanson2017julia} \\
 Rust  & \checkmark & \text{\sffamily X} & \checkmark & \checkmark & \checkmark & \checkmark & MIT & \cite{matsakis2014rust}  \\\bottomrule
    \end{tabular}
    
\end{table}

The paper is structured as follows: Section~\ref{sec:model:problem} introduces the model problem. Section~\ref{sec:overview:languages} briefly overview the languages and libraries and emphasizes the challenges and benefits we experienced during the implementation. Section~\ref{sec:implementation:details} compares programming details. Section~\ref{sec:code:comparison} compares code metrics, \emph{e.g.} the number of lines of codes needed to implement the benchmark. Section~\ref{sec:performance:measurements} compares the performance of the codes on Intel, AMD, and A64FX CPUs.

\section{Model problem}
\label{sec:model:problem}
The one-dimensional heat equation on a 1-D loop (\emph{e.g.}\ limp noodle) $(0 \leq x < L)$ with the length $L$ for all times $t>0$ is described by
\begin{align}
    \frac{\partial u}{\partial t} = \alpha \frac{\partial^2 u}{\partial x^2}, \quad 0 \leq x < L, t> 0 \text{,}
\label{eq:heat}
\end{align}
with $\alpha$ as the material's diffusivity. For the discretization in space, we use
the $N$ grid points $x = \{ x_i = i \cdot h \in \mathbb{R} \; \vert \; i=0,\ldots,N-1 \}$, with the grid spacing $h$ and we use $2^\mathrm{nd}$ order finite differencing.
For the discretization in time, we use the Euler method, \emph{i.e.} 
\begin{align}
u(t+\delta t,x_i) = u(t,x_i) + \delta t \cdot \alpha \frac{u(t,x_{i-1}) - 2 \cdot u(t,x_i) + u(t,x_{i+1})}{2h} 
\text{,}\label{eq:heat:discrete}
\end{align}
with the initial condition $u(0, x_i)  = x_i$.  To model a loop, we use periodic boundary conditions, \emph{i.e.}\ $u(t,x) = u(t,L+x)$.

The parallel algorithm was implemented by having multiple threads of execution each sequentially applying Eq.~\ref{eq:heat:discrete} on a local segment of the grid. We used queues to communicate ghost zones between the segments. We note that for this problem, the queues are single-producer, single-consumer and, therefore, in principle, don't need synchronization (although synchronization to suspend/resume threads seemed to help in some cases).

While this problem is small and extremely simple, it has much in common with many high performance codes that simulate black holes, coastal waves, atmospheres, etc. Block-structured meshes that use finite differences are common and essential for modeling a wide variety of physical systems.

\section{Programming languages and libraries}
\label{sec:overview:languages}
In this section, we comment on the benefits and challenges of each language or framework for the example problem. 
All code is available on GitHub\footnote{\url{https://github.com/diehlpk/async\_heat\_equation}}. Table~\ref{tab:approaches:overview} shows the availability of each approach on various OS, the Licence used, etc.


\subsection{Chapel}
\label{sec:code:chapel}
The Chapel code was one for which we needed to write our own queue. In this case, the full/empty bit synchronization mechanism that the language provides was helpful.
The \texttt{coforall} loop, which assigns a different thread to each iteration, provided a convenient mechanism for launching the outer loop.

Because Chapel is designed to make writing parallel programming easier, we decided also to test a synchronous version of the heat code that did not attempt to segment the array. This code did not scale quite as well as the segmented code that exchanged ghost zones, but the simplicity of the code should make this an appealing option. Though not directly part of the heat equation, we found that Chapel also lacked a built-in way to append to a file. However, opening a file, seeking to the end, and writing is possible. We also add that the support we received from questions asked in the Chapel Gitter was exceptional. We found Chapel to be among the higher performing codes, comparable to Rust or C\texttt{++}. 

\subsection{Charm\texttt{++}}
\label{sec:code:charm}
Charm\texttt{++} derives parallelism through chares that runs on a Processing Element (PE), usually a core of a processor. Multiple chares run asynchronously while communicating using messages. To implement our stencil code, we decomposed the grid into a 1D chare array with neighbors sharing the ghost value every iteration through messages. Send and Receive functions have entry method attribute set as \texttt{expedited}. This allows messages to go directly to the converse scheduler reducing latency further. A reduction at the end triggers the completion of the heat exchange to main-chare.

\subsection{C\texttt{++}\ 17}
\label{sec:code:cpp}
The C\texttt{++}\ 17 code implements a fork and join approach using the \lstinline{std::thread} library. A small queue class was coded by hand, using \lstinline{std::mutex} and \lstinline{std::condition_variable}. Because the queue is single-producer single-consumer, it does not require explicit synchronization. Nevertheless, we synchronize when the queue is empty because that seems slightly faster. C\texttt{++} does not currently provide guarantees of memory safety or data race safety as Rust does, but we note that the language landscape is constantly changing. We expect that ``profiles'' will eventually provide these guarantees for C\texttt{++}.

\subsection{C\texttt{++} Standard Library for Concurrency and Parallelism (HPX)}
\label{sec:code:hpx}
The HPX code was based on the C\texttt{++} code, except that it uses \lstinline{hpx::thread} instead of \lstinline{std::thread}, and uses \lstinline{chanel_spsc} instead of a hand-written queue. The major difference is that HPX uses lightweight threads on top of the operating system threads to avoid context-switching overheads.

\subsection{Go}
\label{sec:code:go}
We use \go{go func} to launch worker threads (goroutines) and buffered channels using \go{make()} to facilitate the exchange of ghost zones. For synchronization of the goroutines, we use \go{sync.WaitGroup} and add threads by calling \go{waitGroup.Add()}, and synchronize the threads by calling \go{waitGroup.Wait()}. As this paper was written, Go was not yet heavily used in scientific computing. \textcolor{black}{At the time of this writing, only \textit{biogo}, an HPC bioinformatics toolkit~\cite{Kortschak2017}, is available.}

\subsection{Julia}
\label{sec:code:julia}
Both Python and Fortran clearly inspire Julia. It is a good choice for Fortran programmers who want to get into scripting, as it will offer some familiarity in using one as the default start for array indexes (instead of zero) and its use of end to mark the end of a block.

In our Julia code, we implemented our own queue. Since Julia does not support classes directly (though it has structs), we found it convenient to use arrays. For parallelism, we used Julia's \texttt{Thread.@threads for} loop macro.

\subsection{Python}
\label{sec:code:python}
For the grid segments, we use \python{NumPy}~\cite{harris2020array}, which gives us SIMD out of the box. For concurrency, we use \python{from threading import Thread} to launch the worker for each data segment, and we use \python{from queue import Queue} to synchronize the exchange of ghost nodes. The queue is, therefore, synchronized and possibly has lower performance than it could be. We did attempt to create a queue class by hand, as we did with C\texttt{++}, but it did not perform better than the built-in \texttt{queue.Queue} object. Unsurprisingly, perhaps, Python is the slowest code of the set we are considering in this paper, despite using NumPy. This is undoubtedly because it is an interpreted language. While Julia is faster than Python, the startup time seems to be a bit longer (especially with option \texttt{-O3}).


\subsection{Rust}
\label{sec:code:rust}
We use \rust{std::thread::scope} to launch worker threads, and non-blocking channels from \rust{std::sync::mpsc} to facilitate the exchange of ghost zones.
We avoided using \rust{unsafe}, working only in the safe subset of Rust. In addition to the regular implementation of the heat code, we have a version with SIMD vectorization using the unstable \rust{portable_simd} feature. This improves performance over the non-vectorized version with a few threads, but the improvement diminishes as the number of threads increases. We considered only the non-SIMD version in our final measurements to remain consistent with the other languages. Rust is not yet widely adopted in scientific computing. However, we found two such scientific codes that are available: \textit{Lumol}\footnote{\url{https://lumol.org/}}, a molecular simulation engine; and \textit{Rust-Bio}~\cite{koster2016rust}, a bioinformatics library. Because of its guarantees concerning data race conditions and memory access, as well as its high performance, Rust is a potentially good choice for new scientific programming projects. However, Rust has vastly different syntax and semantics than more traditional languages like C\texttt{++}, Java, and Python, all of which may make for a steep learning curve.

\subsection{Swift}
\label{sec:code:swift}
We had to use \swift{UnsafeMutableBufferPointer<Double>} to avoid unnecessary calls of \swift{await} for accessing the elements of arrays. These buffers allow explicit vectorization on newer x86 and Apple Silicon. See, for example, \swift{addingProduct}. However, we could not measure a large improvement using these functions. For concurrency, we use \swift{await with TaskGroup\{ body: \{ group in\}\}} to launch chunks of works on each thread and \swift{for wait _ in group\{\}}. Swift claims to be safe by design and produces software that runs lightning-fast. Unfortunately, we had to disable the safety feature to get a performant code. It may be that simulating the heat equation is not a good fit for Swift's design. We have not yet found any scientific simulation codes built on Swift. Python and Julia share similar patterns. Chapel is somehow unique due to its compactness.
 
\subsection{Java}
\label{sec:code:java}

The Java code was of intermediate performance. This was somewhat expected since it does not target high performance computing. However, it is not unknown either. Java has several thread-safe queue objects. We used an \texttt{ArrayBlockingQueue} for the ghost zone exchange, giving it a size equal to the number of threads (approximately twice as much as it could theoretically use). It also has several threading systems to choose from. In this case, we chose the FixedPoolThread, which matches our consistent and uniform workload..

Overviews of Java in HPC are available here~\cite{10.1145/1596655.1596661,amedro2008current}. One virtue that Java continues to have is simplicity. It was designed for ``average programmers'' and does not have a lot of syntax a new programmer needs to learn (though it becomes a little more complex with each version). It's choice to borrow much of C\texttt{++}'s constructs makes it feel familiar from its first use.

\section{\textcolor{black}{Remarks on the implementation details}}
\label{sec:implementation:details}
This section focuses on the implementation details and showcases similarities and differences. One aspect is the safety, \emph{e.g.}\ memory, included in the language. Rust and Swift have implemented some sanity checks, for example, for memory access. For Rust, we could avoid using \rust{unsafe} and worked only in the safe subset of Rust. However, for Swift, we had to use \swift{UnsafeMutableBufferPointer} to get the reported performance. Another aspect is the usage of SIMD for scientific computations. Here, some of the approaches, like Python, allowed SIMD as the default for NumPy. Other approaches, like Rust and Swift, provide SIMD as a language feature, while others rely on external libraries, like C\texttt{++}, Charm\texttt{++}, HPX, or Julia.
Measuring the elapsed time and parsing command line options was decent in all the languages. Creating files and appending the performance data to the file was more accessible in some than others. For example, Chapel lacks a built-in way to append to a file. Handling files in Swift was rather complex; we just piped it to the console. For the synchronization for the ghost elements, we decided to rely on Queues for our code. Some approaches, like Python, Rust, and Go, provide queues or equivalent mechanisms. For other approaches, we had to implement the queue using features included in the language. Next to these features, we identified similar code patterns. For example, the Swift code (\swift{TaskGroup})and Go code (\go{WaitGroup}) uses similar semantics to group the threads and synchronize them. HPX, C\texttt{++}, and Rust use fewer abstraction layers and act on the bare thread libraries.

\section{Comparison of the asynchrony}
\label{sec:code:comparison}

\begin{figure}[tb]
    \centering
    \subfloat[\label{fig:line:of:codes}]{

 \resizebox{0.45\textwidth}{!}
        {
  \begin{tikzpicture}
 \begin{axis}[
    xbar=12pt,
    xmin=0,xmax=220,
    ytick=data,
    enlarge y limits={abs=1cm},
    symbolic y coords={Python,Swift,HPX,Julia,Go,Rust,Chapel,Charm\texttt{++},C\texttt{++}\ 17,Java},
    bar width = 10pt,
    xlabel= Lines of code (LOC), 
    ytick align=outside, 
    ytick pos=left,
    major x tick style ={ transparent},
    legend style={at={(0.04,0.96)},anchor=north west, font=\footnotesize, legend cell align=left},
    xmajorgrids=true
        ]    
    \addplot[xbar,fill=cadetgrey!20, area legend] coordinates {
        (66,Python)
        (111,Swift)
        (109,HPX)
        (86,Julia)
        (120,Go)
        (134,Rust)
        (44,Chapel)
        (160,Charm\texttt{++})
        (139,C\texttt{++}\ 17)
        (152,Java)
        };
\end{axis}
 
\end{tikzpicture}
        }
    
    }
\hfill
    \subfloat[\label{fig:two:dim:plot} ]{

 \resizebox{0.45\textwidth}{!}
        {
  \begin{tikzpicture}
    \draw[help lines, color=gray!30, dashed] (-2.9,-2.9) grid (2.9,2.9);
    \draw[<->,thick,cadetgrey] (-3,0)--(3,0) node[right]{Difficult};
    \draw[<->,thick,cadetgrey] (0,-3)--(0,3.1) node[above,cadetgrey]{Fast};
    \node[left,cadetgrey] at (-3,0) {Easy};
    \node[below,cadetgrey] at (0,-3) {Slow};
    \draw[fill=black] (-2.470588235,-3) circle [radius=0.05] node[right] {Python};
    \draw[fill=black] (1.058823529,1.345362794) circle [radius=0.05] node[right] {Go};
    \draw[fill=black] (-1.235294118,1.802650018) circle [radius=0.05] node[left] {Julia};
    \draw[fill=black] (1.323529412,2.990484116) circle [radius=0.05] node[right] {Rust};
    \draw[fill=black] (-3,2.967574361) circle [radius=0.05] node[right] {Chapel};
    \draw[fill=black] (1.235294118,2.990096568) circle [radius=0.05] node[above] {C\texttt{++}\ 17};
     \draw[fill=black] (-0.088235294,3) circle [radius=0.05] node[left] {HPX};
     \draw[fill=black] (0,2.970081617) circle [radius=0.05] node[below] {Charm\texttt{++}};
    \draw[fill=black] (-0.088235294,2.272308495) circle [radius=0.05] node[left] {Swift};
     \draw[fill=black] (3,2.654365913) circle [radius=0.05] node[below] {Java};
    \end{tikzpicture}
        }
    
    }

    \caption{Software engineering metrics: \protect\subref{fig:line:of:codes} Lines of codes for all implementations. The numbers were determined with the Linux tool \textit{cloc} and   \protect\subref{fig:two:dim:plot} Two-dimensional classification using the computational time and the COCOMO model.}
    
\end{figure}
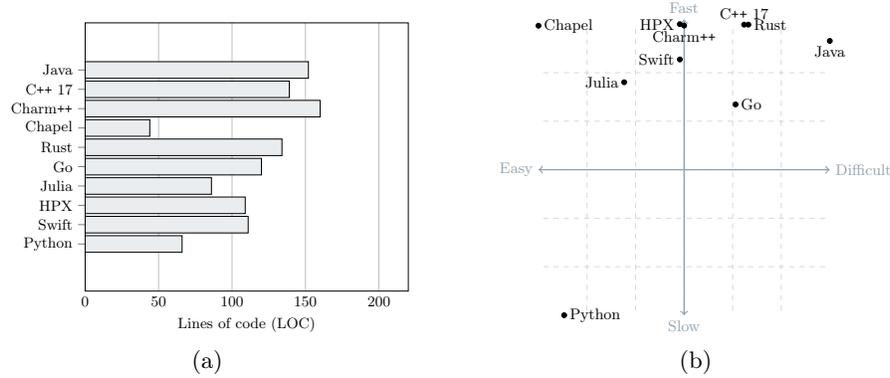

Figure~\ref{fig:line:of:codes} shows the lines of code (LOC) for each approach. The numbers were determined with the Linux tool \lstinline[language=bash]{cloc}\footnote{\url{https://github.com/AlDanial/cloc}} and exclude comments. Here, Chapel has the least amount of lines of code due to its high-level language design. Second, is Python and third is Julia. Next, Swift, HPX, and Go have similar lines of code. Rust and C\texttt{++} are close. Charm\texttt{++} and Java have the most lines of code. While we tried to be reasonably uniform in our coding methodology, we note that some codes include a hand-written queue and some do not. Regardless, these numbers might differ if a different person coded the problem.

Even so, lines of code are just one metric and do not tell the whole story about the code's complexity. Inspired by~\cite{9484790}, we use a scale to determine how complicated it was to develop the code and relate it to the performance of the code, see Figure~\ref{fig:two:dim:plot}. For the scale from slow to fast, we use the computational time on Intel CPUs in Figure~\ref{fig:performance:intel}. We use the average 
\begin{align}
    T_\text{average}(\text{approach}):= \sfrac{\left(T_2(\text{approach}) + T_{20}(\text{approach}) + T_{40}(\text{approach})\right)}{3}
\end{align}
of the computational time on two cores $T_2$, on \num{20} cores $T_{20}$, and \num{40} cores $T_{40}$. We arrange these averages linearly on the interval $[-1,1]$ on the $y$ scale. Quantifying the performance based on computational time was quite objective. However, quantifying how easy or hard it was to implement the code in various approaches might be subjective. Therefore, we agreed on using the \textbf{Co}nstructive \textbf{Co}st \textbf{Mo}del (COCOMO)~\cite{stutzke1997software,barry1981software}. Note that the COCOMO model is a general model without any specialization in parallel programming. There has been much discussion in the HPC community about the model, however, the HPC community did not invent a specialized model yet. For example, the COCOMO\ II model was used in~\cite{miller2018applicability} to analyze its cost parameters for the investigated parallelization projects with OpenACC on NVIDIA GPUs. Some overview of performance metrics in HPX is given in~\cite{kepner2004high}. Until we get a specialized model for parallel computing or HPC, the COCOMO model is a reasonable candidate.
We use the tool \textit{Sloc Cloc and Code} (scc)\footnote{\url{https://github.com/boyter/scc}} to get the COCOMO metrics for each approach. We use the Estimated Schedule Effort using the organic option to estimate how long it would take to implement the code in a month. We arranged these values on linearly on the interval $[-1,1]$ on the $x$ scale. Figure~\ref{fig:two:dim:plot} shows the classification using this metric. The first cluster is Python, Julia, and Chapel. These three languages were classified as easier to implement for this specific application.

One possible cause of Python's slowness might be the queue implementation. However, our own queue implementation did not help. The second cluster is Swift, Charm\texttt{++}, and HPX. Where HPX and Charm\texttt{++} have a slightly better performance. The third cluster is C\texttt{++}\ 17, Rust, and go. Here the performance of C\texttt{++}\ 17 and Rust were comparable. Interestingly, Java had a decent performance and was classified as the most difficult. Mostly, we assume, due to much boilerplate code produced by Java. Subjectively, we would identify Java as one of the more accessible codes to implement, so the COCOMO classification is not perfect. However, it is a place to start in assessing various programming languages.

\section{Performance measurements}
\label{sec:performance:measurements}
In this section, we compare the performance of the approaches on the following CPUs: ARM A64FX, AMD EPYC\textsuperscript{\texttrademark} 7543, and Intel\textsuperscript{\textregistered} Xeon\textsuperscript{\textregistered} Gold 6140, respectively. Table~\ref{tab:lib:compilers} shows the versions and dependencies of the compilers or interpreters. We report the cells processed per second for all methods. We know that FLOPS would be a better quantity. However, not all the approaches provided a convenient interface to Papi.
Figure~\ref{fig:performance} shows the computation time for all approaches on all three architectures. Table~\ref{tab:performance:r2} shows the $R^2$ correlation for all line fits computed using Python NumPy.

\begin{table}[tb]
    \centering
      \caption{Software versions and dependencies. The nightly build from April 23 was used for Rust, since we needed SIMD support. HPX was built using boost 1.78, hwloc 2.8.0, and jemalloc 5.3. If needed, CMake 3.20 was used for building.}
    \label{tab:lib:compilers}
    \begin{tabular}{llllll}\toprule
     Python 3.10 & Julia 1.8.5 & Chapel 1.30  & Go 1.20.3  & Charm\texttt{++} \textit{6f7f105b}   \\
     Swift 5.8  & HPX 1.8.1 & OpenJDK 20 &  gcc 11.2.1 & Rust nightly (April 23)  \\\bottomrule
    \end{tabular}
  
\end{table}

\subsection{Intel}
Figure~\ref{fig:performance:intel} shows the performance on Intel\textsuperscript{\textregistered}Xeon\textsuperscript{\textregistered}Gold 6148 Skylake. Here, Python was the slowest approach. Swift and Julia are comparable. For larger than \num{10} threads Go behaves slightly better than Swift and Julia. For smaller core counts up to eight cores, the remaining approaches behave similarly. However, Chapel gets slower for higher node counts. For Rust, Charm\texttt{++}, and HPX the performance is comparable. HPX is for larger node counts the fastest, but has a high variance, see $R^2$ in Table~\ref{tab:performance:r2}.

\subsection{AMD}
Figure~\ref{fig:performance:amd} shows the performance on AMD EPYC 7H12. Python is the slowest approach. After that, Swift and Julia are comparable for smaller core counts, but after that, Swift gets slower. Go is faster, but had a high variation. Chapel and Rust are comparable. Charm\texttt{++} and C\texttt{++} are close. HPX is comparable, but with a high variance, see $R^2$ in Table~\ref{tab:performance:r2}.

\subsection{A64FX}
\begin{figure}[tb]
    \centering
      \subfloat[Intel\textsuperscript{\textregistered}Xeon\textsuperscript{\textregistered}Gold 6148 Skylake\label{fig:performance:intel}]{
      \resizebox{0.4\textwidth}{!}
        {

        \includegraphics[width=\textwidth]{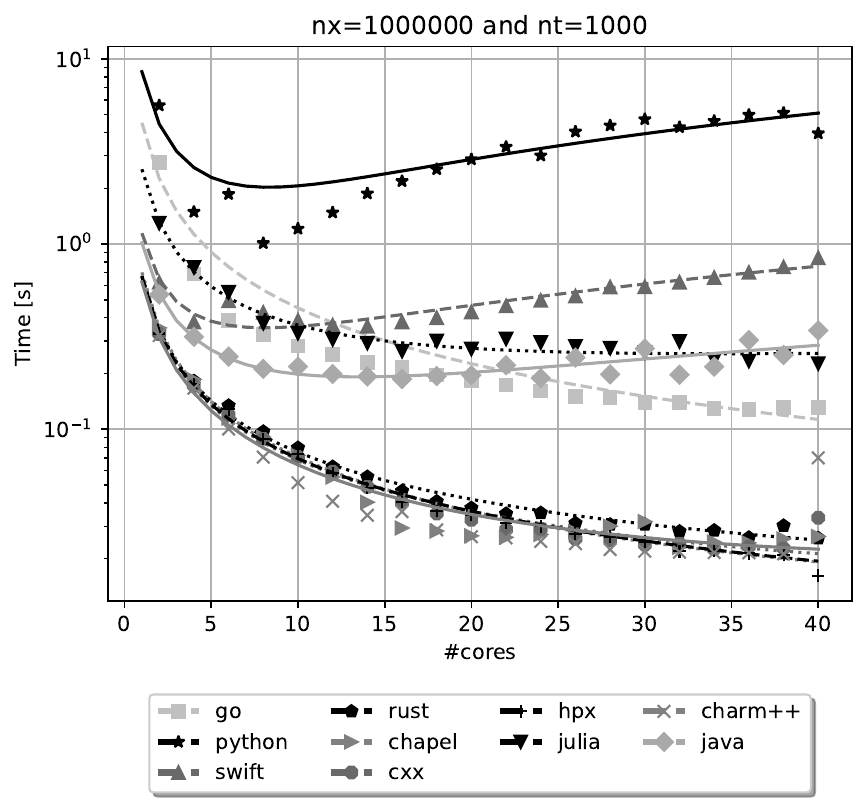}
}
}
      \subfloat[AMD EPYC 7H12\label{fig:performance:amd}]{

      \resizebox{0.4\textwidth}{!}
        {

   \includegraphics[width=\textwidth]{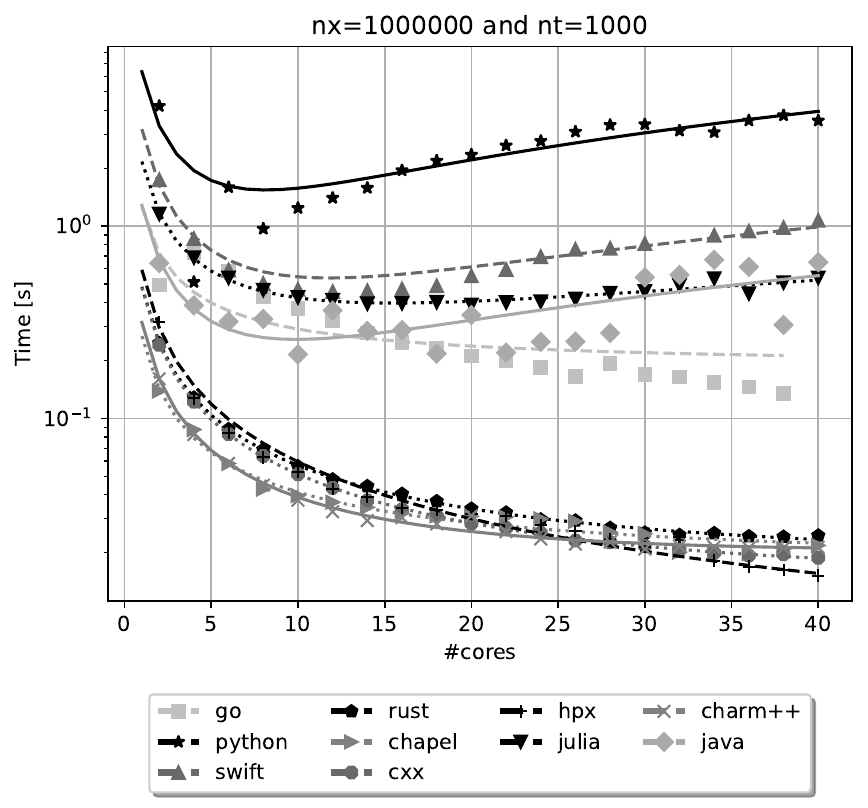}

    }
}

      \subfloat[Arm 64FX \label{fig:performance:arm} ]{
  
            \resizebox{0.4\textwidth}{!}
        {
 \includegraphics[width=\textwidth]{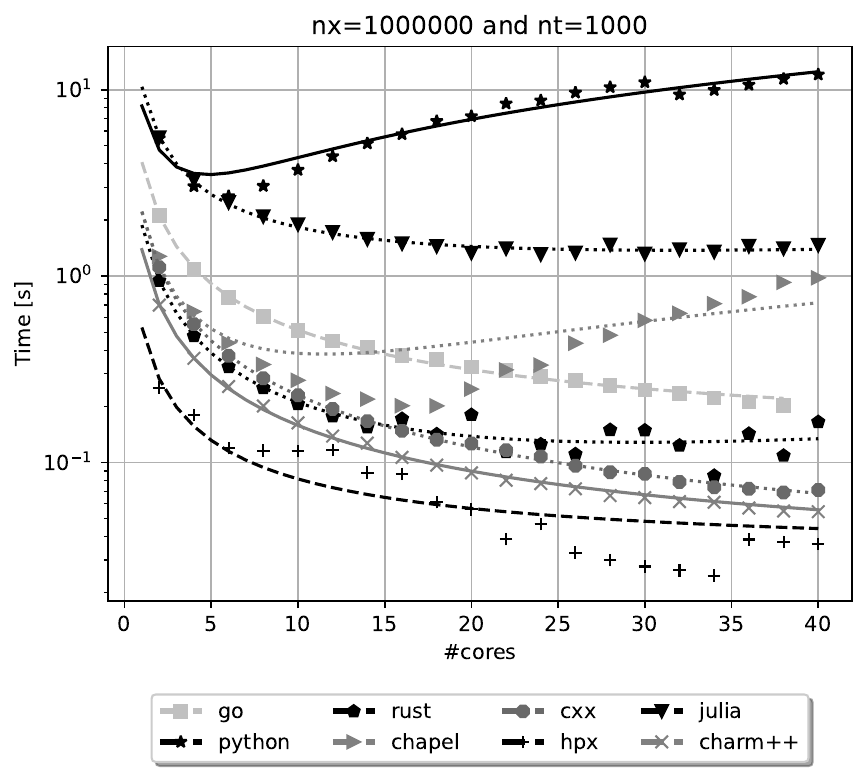}
    }
      }
    \caption{The obtained performance for three different architectures: Intel \protect\subref{fig:performance:intel}, AMD \protect\subref{fig:performance:amd}, and A64FX \protect\subref{fig:performance:arm}. The baseline was \num{1000000} discrete nodes and \num{1000} time steps. Swift is missing on A64FX, since no package was provided for Rocky Linux. The lines are the curve fits obtained with \lstinline[language=python]{curve_fit} from Python SciPy.}
    \label{fig:performance}
\end{figure}

Note that Ookami uses \textit{Rocky Linux} and no Swift package was provided for this OS. Rocky Linux was developed to be a binary-compatible with Red Hat Enterprise Linux, however, Red Hat Universal package provided by Swift had some library version mismatches. Figure~\ref{fig:performance:arm} on A64FX. Here, Python was the slowest, as on all architectures. Chapel behaves differently on A64FX and is the second slowest. Go is in the middle as before. Rust and C\texttt{++} are comparable up to \num{16} cores, and C\texttt{++} outperforms Rust for larger node counts. Charm\texttt{++} is the second fastest. HPX is the fastest system on A64FX, however with a large variation, see the $R^2$ in Table~\ref{tab:performance:r2}. We want to emphasize that some of the approaches might not optimize for A64FX yet and just support the architecture, since they might not target it.

\newcommand{\roundnumber}[1]{\num[round-mode=places,round-precision=2]{#1}}
\begin{table}[tb]
    \centering
    \caption{$R^2$ correlation of the fit of the measured data points for all approaches and architectures, computed using Python NumPy.}
    \begin{tabular}{l|cccccccccc} \toprule
     Architecture    & C\texttt{++} & Charm\texttt{++} & Chapel & Rust & Go & Julia & HPX & Swift & Python & Java  \\\midrule
    Intel\textsuperscript{\textregistered}Xeon\textsuperscript{\textregistered}Gold 6148  &     \roundnumber{0.4895839842408277} & \roundnumber{0.36373253705547925} & \roundnumber{0.4486974703706921} & \roundnumber{0.5175252397325332} & \roundnumber{0.28099783457284816} & \roundnumber{0.41394743270443446} & \roundnumber{0.5231489248583688} & \roundnumber{0.5636319573657262} & \roundnumber{0.43045131026500316} & \roundnumber{0.028005174773163444}   \\
    AMD\textsuperscript{\texttrademark} EPYC\textsuperscript{\texttrademark} 7H12 &   \roundnumber{0.4836434576286168} & \roundnumber{0.45151166336552506} & \roundnumber{0.5290852650106667} & \roundnumber{0.4891098256032546} & \roundnumber{0.7470580460870956} & \roundnumber{0.12381192693325543} & \roundnumber{0.41938407375070574} & \roundnumber{0.023027154570507576} & \roundnumber{0.45678496199021695} & \roundnumber{0.12380846694614872} \\
    Arm A64FX & \roundnumber{0.49440357979739796} & \roundnumber{0.5191769223313201} & \roundnumber{0.0836619759029489} & \roundnumber{0.4027935404454299} & \roundnumber{0.5195310218991788} & \roundnumber{0.4203967297689673} & \roundnumber{0.7288119902019908} & -- & \roundnumber{0.8991221497169645} & \roundnumber{0.32449190246502035} \\\bottomrule
    \end{tabular}

    \label{tab:performance:r2}
\end{table}

\section{Conclusion}
\label{sec:outlook}

There are an increasing number of viable platforms for creating high performance programs, each with its own benefits and limitations. While these platforms can make the creation of high performance applications easier, learning to use a new language requires a shift in mindset and a significant learning effort. Even small differences, such as starting arrays with 1 instead of 0, can create confusion.

Often, knowledge of one programming language can mislead a programmer into thinking they understand another.
The Chapel ``class'' for example did not behave exactly as we expected from C\texttt{++} and Java.

Ideally, any new language should have tutorials that help new users transition from a familiar platform such as C\texttt{++} or Python to something new. This allows experienced programmers to come up to speed more quickly without having to slog through fairly familiar territory such as what control flow is or how to declare a variable.

Python was the slowest of the set we compared, despite the use of NumPy. This is not surprising given that it is interpreted and not built for performance. However, because of the language's simplicity and short REPL time, it may be a good choice for prototyping or situations where performance is not critical. It is clearly one of the most popular programming languages available, according to the 2023 Tiobe index\footnote{\url{https://doi.org/10.1038/s41586-020-2649-2}}.

Java, Go, Swift, and Julia were the intermediate performers in our experiments. We list them in the order of their popularity in the Tiobe Index. We note that we have the least expertise in Go, Swift, and Julia relative to the others.

The higher performing platforms were C\texttt{++}, Rust, Chapel, Charm\texttt{++}, and HPX. Again, we list in the order of the Tiobe Index. We note that of this set, HPX is not a language but a runtime library for C\texttt{++}, and Charm\texttt{++} is a declaration language designed to be used in conjunction with C\texttt{++} rather than a full-blown language.

We will not name a winner with respect to speed. The higher performing platforms were mostly similar in what they achieved. The tests in this paper are dependent on the hardware, the version of the interpreters and compilers, the particular problem chosen, the amount of effort applied, and our level of expertise (which varied by platform).

One solution does not fit all. Despite being the slowest, Python will undoubtedly continue to be one of the most popular. C\texttt{++} will likely remain ``everyone's second choice.'' Rust's innovative ideas will make it ideally suited for many applications. Chapel and Julia make many HPC tasks easier.

\subsection*{Future Work}
Since this paper only considers a stencil-based one-dimensional code, other numerical applications would be needed for a more comprehensive comparison. Various collections of high performance codes and parallel algorithms exist, many of which have significant overlap in the kinds of things they do. Can a good representative subset be identified? One significant question would be to investigate distributed programming for some of these languages and platforms. For Python and Rust, MPI bindings are available. However, Chapel, Charm\texttt{++}, Julia, and HPX provide higher abstraction levels for communication. What benefits do these approaches have for the programmer? For high performance programming in general? \textcolor{black}{Another aspect to explore is the GPU support using native languages, like CUDA or RocM; or abstraction layers, like Kokkos or SYCL.}

\section*{Supplementary materials}
{\footnotesize
The code for all examples is available on GitHub\textsuperscript{\textregistered}\footnote{\url{https://github.com/diehlpk/async\_heat\_equation}} or Zenodo\textsuperscript{\texttrademark}~\cite{patrick_diehl_2023_7942453}, respectively. A Docker image to compile/run all examples is available here\footnote{\url{https://hub.docker.com/r/diehlpk/monte-carlo-codes}}.}

\section*{Acknowledgments}
{\footnotesize
The authors would like to thank Stony Brook Research Computing and Cyberinfrastructure, and the Institute for Advanced Computational Science at Stony Brook University for access to the innovative high-performance Ookami computing system, which was made possible by a \$5M National Science Foundation grant (\#1927880). We thank Steve Canon and Nick Everitt for their remarks on the Swift code and Brad Chamberlain and Jeremiah Corrado for their remarks on the Chapel code.  }
%
%
%
\bibliographystyle{splncs04}
\bibliography{ref}

\begin{thebibliography}{10}
\providecommand{\url}[1]{\texttt{#1}}
\providecommand{\urlprefix}{URL }
\providecommand{\doi}[1]{https://doi.org/#1}

\bibitem{amedro2008current}
Amedro, B., et~al.: Current state of Java for HPC. Ph.D. thesis, INRIA (2008)

\bibitem{arnold2005java}
Arnold, K., Gosling, J., Holmes, D.: The Java programming language. Addison
  Wesley Professional (2005)

\bibitem{barry1981software}
Barry, B., et~al.: Software engineering economics. New York  \textbf{197}
  (1981)

\bibitem{bennett2015asc}
Bennett, J., et~al.: {ASC ATDM level 2 milestone\# 5325: Asynchronous many-task
  runtime system analysis and assessment for next generation platforms}.
  SAND2015-8312  (2015)

\bibitem{bezanson2017julia}
Bezanson, J., et~al.: Julia: A fresh approach to numerical computing. SIAM
  review  \textbf{59}(1),  65--98 (2017)

\bibitem{chamberlain2015chapel}
Chamberlain, B.L., Deitz, S., Hribar, M.B., Wong, W.: Chapel. Programming
  Models for Parallel Computing pp. 129--159 (2015)

\bibitem{chamberlain2007parallel}
Chamberlain, B.L., et~al.: Parallel programmability and the chapel language.
  The International Journal of High Performance Computing Applications
  \textbf{21}(3),  291--312 (2007)

\bibitem{patrick_diehl_2023_7942453}
Diehl, P., et~al.: {Benchmarking the Parallel 1D Heat Equation Solver in
  Chapel, Charm\texttt{++}, C\texttt{++}, HPX, Go, Julia, Python, Rust, Swift,
  and Java} (2023). \doi{10.5281/zenodo.7942453},
  \url{https://doi.org/10.5281/zenodo.7942453}

\bibitem{godoy2023evaluating}
Godoy, W.F., et~al.: Evaluating performance and portability of high-level
  programming models: Julia, python/numba, and kokkos on exascale nodes (2023)

\bibitem{harris2020array}
Harris, C.R., et~al.: Array programming with {NumPy}. Nature
  \textbf{585}(7825),  357--362 (Sep 2020)

\bibitem{kaiser2020hpx}
Kaiser, H., et~al.: {HPX-the C\texttt{++} standard library for parallelism and
  concurrency}. Journal of Open Source Software  \textbf{5}(53), ~2352 (2020)

\bibitem{kale1993charm++}
Kale, L.V., Krishnan, S.: {Charm\texttt{++} a portable concurrent object
  oriented system based on C\texttt{++}}. In: Proceedings of the eighth annual
  conference on Object-oriented programming systems, languages, and
  applications. pp. 91--108 (1993)

\bibitem{kepner2004high}
Kepner, J.: High performance computing productivity model synthesis. The
  International Journal of High Performance Computing Applications
  \textbf{18}(4),  505--516 (2004)

\bibitem{Kortschak2017}
Kortschak, R.D., et~al.: bíogo: a simple high-performance bioinformatics
  toolkit for the go language. Journal of Open Source Software  \textbf{2}(10),
  ~167 (2017)

\bibitem{koster2016rust}
K{\"o}ster, J.: Rust-bio: a fast and safe bioinformatics library.
  Bioinformatics  \textbf{32}(3),  444--446 (2016)

\bibitem{matsakis2014rust}
Matsakis, N.D., Klock~II, F.S.: The rust language. In: ACM SIGAda Ada Letters.
  vol.~34, pp. 103--104. ACM (2014)

\bibitem{miller2018applicability}
Miller, J., et~al.: {Applicability of the software cost model COCOMO II to HPC
  projects}. International Journal of Computational Science and Engineering
  \textbf{17}(3),  283--296 (2018)

\bibitem{9484790}
Pennycook, S.J., et~al.: Navigating performance, portability, and productivity.
  Computing in Science \& Engineering  \textbf{23}(5),  28--38 (2021)

\bibitem{stutzke1997software}
Stutzke, R.D., Crosstalk, M.: Software estimating technology: A survey. Los.
  Alamitos, CA: IEEE Computer Society Press (1997)

\bibitem{10.1145/1596655.1596661}
Taboada, G.L., et~al.: Java for high performance computing: Assessment of
  current research and practice. In: Proceedings of the 7th International
  Conference on Principles and Practice of Programming in Java. p. 30–39
  (2009)

\bibitem{10.5555/1593511}
Van~Rossum, G., Drake, F.L.: Python 3 Reference Manual. CreateSpace, Scotts
  Valley, CA (2009)

\bibitem{10.1007/978-3-319-41321-1_17}
Van~der Wijngaart, R.F., et~al.: Comparing runtime systems with exascale
  ambitions using the parallel research kernels. In: High Performance
  Computing. pp. 321--339. Springer, Cham (2016)

\end{thebibliography}

\end{document}